\documentclass{article}
\usepackage{spconf,amsmath,graphicx}
\usepackage{multirow}

\title{Hybrid Attention-based Encoder-decoder Model for Efficient Language Model Adaptation}
\name{Shaoshi Ling, Guoli Ye, Rui Zhao, Yifan Gong}
\address{Microsoft Cloud and AI}
\begin{document}
%
\maketitle
\begin{abstract}
The attention-based encoder-decoder (AED) speech recognition model has been widely successful in recent years. However, the joint optimization of acoustic model and language model in end-to-end manner has created challenges for text adaptation. In particular, effective, quick and inexpensive adaptation with text input has become a primary concern for deploying AED systems in the industry. To address this issue, we propose a novel model, the hybrid attention-based encoder-decoder (HAED) speech recognition model that preserves the modularity of conventional hybrid automatic speech recognition systems. Our HAED model separates the acoustic and language models, allowing for the use of conventional text-based language model adaptation techniques. We demonstrate that the proposed HAED model yields 23\% relative Word Error Rate (WER) improvements when out-of-domain text data is used for language model adaptation, with only a minor degradation in WER on a general test set compared with the conventional AED model. 
\end{abstract}
\section{Introduction}
\label{sec:intro}

The attention-based encoder-decoder model (AED) \cite{chorowski2014end, chan2015listen, bahdanau2016end, kim2017joint, chiu2018state} has become increasingly popular in both academic and industry for end-to-end (E2E) automatic speech recognition (ASR) due to its highly competitive accuracy on a variety of tasks. However, a major drawback of AED models is their limited ability of adaptation using only text data. Unlike traditional hybrid models, AED models directly learn the mapping from the input features to output labels within a single network with no separated acoustic model (AM) or language model (LM). The encoder maps the input features into high-level representations. The decoder functions as a language model by taking the previously predicted tokens as input. However, it is not a true language model as it is also involved in cross-attention computation and then it will be combined with the weighted-sum of representations from the encoder to generate posteriors over the vocabulary. Adapting the decoder network using text-only data is not as straightforward or effective as adapting the LM in hybrid systems. Normally, paired audio and text data are required to adapt an AED model. Unfortunately, collecting labeled audio data is both time-consuming and expensive.

Numerous studies have explored the adaptation of the AED model using external text-only data. One straightforward approach is to use artificial audio generated from text data instead of collecting real audio.  This can be achieved through either a neural text to speech (TTS) model \cite{li2019semi, sim2019personalization} or a spliced data method \cite{zhao2021addressing}. The AED model could then be finetuned with artificial paired audio and text data. However, a major drawback of these methods is their high computational cost. TTS-based methods require significant time to generate audio and also dilute the information present in real speech training data, including natural speech phenomena such as disfluency, accent, dialect, and acoustic environment, resulting in an increase in WER. Meanwhile, spliced data methods may produce audio that is significantly different from real-world scenarios, especially for unseen domains, making it less effective for training the AED model. Furthermore, since both acoustic and language components need to be adapted with the above methods, the training cost is high, limiting the application in scenarios that require rapid adaptation.

A different type of text-only adaptation methods is LM fusion, such as shallow fusion \cite{gulcehre2015using, toshniwal2018comparison} where an external LM trained on target-domain text is incorporated during the decoding. However, this approach could be problematic because the AED model already contains an internal LM, and simply adding an external LM may not be mathematically sound. To overcome this issue, researchers have proposed density ratio \cite{mcdermott2019density, zeyer2021librispeech} and internal LM estimation based approaches \cite{meng2021internal, meng2021internal2, zeineldeen2021investigating, liu2022internal}, which aim to remove the influence of the internal LM contained in the AED model. However, these methods can be sensitive to the interpolation weight of the LM for different tasks and require careful tuning using development data to achieve optimal results \cite{meng2021internal2}. Furthermore, they can increase both the computational cost and model footprint \cite{tsunoo2022residual}, which may not be efficient in a production system.

The paper proposes a novel AED architecture called {\bf hybrid attention-based encoder-decoder} (HAED) which retains the modularity of conventional hybrid automatic speech recognition systems. During training, the decoder is optimized to function as a standard neural language model. As a result, various conventional language model adaptation techniques \cite{bellegarda2004statistical, chen2015recurrent} could be applied to the decoder in HAED for text-only adaptation. The objective is to separate the fusion of AM and LM in E2E models, making language model adaptation and customization more efficient. This effect is similar to the impact of the language model in the conventional hybrid ASR system. Experimental results demonstrate that on adaptation sets, the word error rate (WER) after the adaptation of HAED is reduced by 23\% relative to the baseline model.

\section{Related work}
\subsection{Factorized transducer and AED}
For transducer model \cite{graves2012sequence}, HAT and its variant \cite{variani2020hybrid, meng2023modular} estimate the internal LM using the prediction network and demonstrated that internal LM is proportional to the prediction network when without the effect of the encoder. The factorized transducer (FNT) work \cite{chen2022factorized, zhao2023fast, gong2023longfnt, levit2023external, gong2024advanced} and the MHAT work \cite{meng2023modular} go beyond HAT by factoring the prediction network into the blank and vocabulary prediction, thereby allowing the vocabulary prediction to function as an independent LM. JEIT \cite{meng2023jeit} further optimized the model with the internal language model training. For the AED model, factorized AED \cite{gong2023factorized} proposed a model whose cross attention takes posterior probability as input to achieve factorization. And in RILM \cite{deng2023adaptable} and Decoupled AED \cite{deng2024decoupled}, the cross-attention modules of the Transformer decoder are decoupled from the self-attention modules to achieve flexible domain adaptation. In our HAED method, we further eliminate the dependency on previous tokens in the cross attention modules to achieve better stand-alone LM capabilities of the decoder in AED model.


\subsection{Attention-based Encoder-decoder}
The AED model calculates the probability using the following equation:
\begin{equation}\label{probability}
  \begin{aligned}
P(y|x)=\prod_uP(y_u|x,y_{0:u-1})
\end{aligned}
\end{equation}
where $u$ represents the output label index, x is the input feature sequence, and y is the output label sequence. The training objective is to minimize $-lnP(y|x)$. The encoder in the model encodes the features sequence $x$ to representations $h$. During each decoder step u, an attention mechanism is used to calculate attention weights by:
\begin{equation}\label{attention}
  \begin{aligned}
\alpha_{u,t}=Softmax(Attention(d_{u-1}, h)))
\end{aligned}
\end{equation}
where $d_{u-1}$ is the hidden states at step $u-1$ and $\alpha_{u,t}$ is the attention score at step $u$ over acoustic time index $t$.

The attention context vector $c_u$ is then computed as weighted sum over the encoder hidden states $h$ as below:
\begin{equation}\label{context vector}
  \begin{aligned}
c_u = \sum _{t=1}^{T}\alpha_{u,t}h_t
\end{aligned}
\end{equation}
The decode sates $d_u$ is computed as:
\begin{equation}\label{decoder}
  \begin{aligned}
d_u = Decoder(d_{u-1}, y_{u-1}, c_{u-1})
\end{aligned}
\end{equation}
Finally, the output probability for label $y_u$ is computed as:
\begin{equation}\label{output}
  \begin{aligned}
P(y_u|x, y_{0:u-1}) = Softmax(MLP(d_u))
\end{aligned}
\end{equation}

In the training recipe introduced by \cite{kim2017joint}, the AED model is also optimized together with a CTC model in a multi-task learning framework where CTC and AED model share the same encoder. Such a training strategy can greatly improve the convergence. Our model below follows this recipe.

\begin{figure}[t]
  \centering
  \includegraphics[width=3cm]{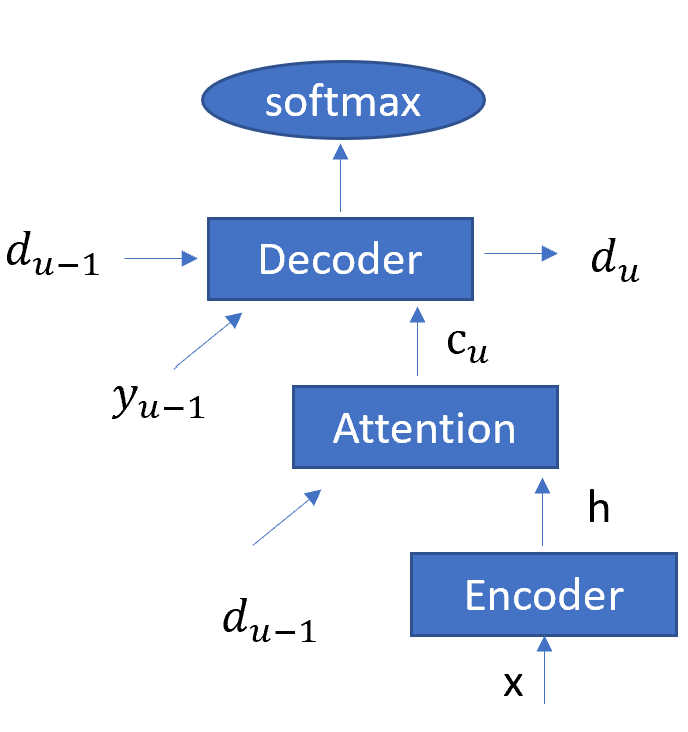}
  \includegraphics[width=4cm]{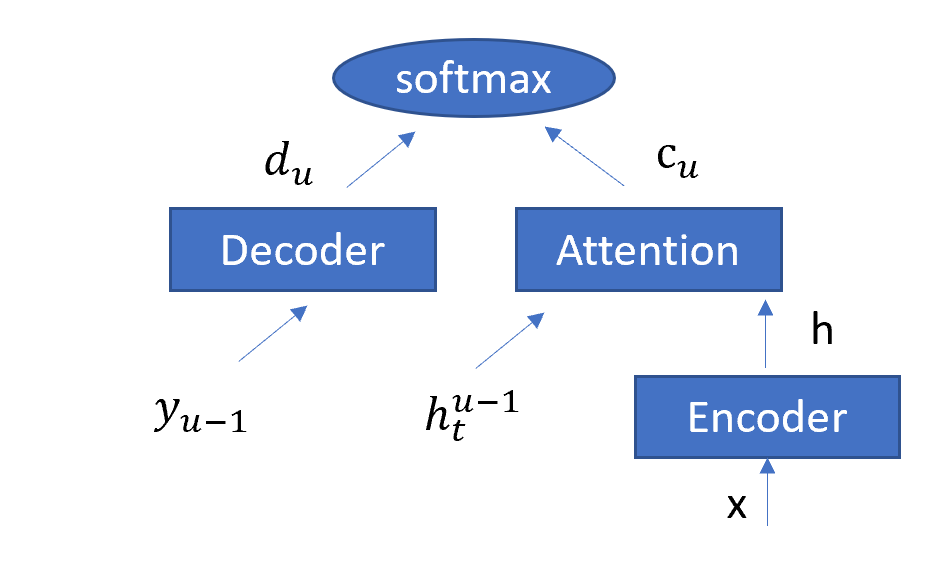}
  \caption{The architecture comparison between AED (left) and HAED (right). HAED model allows its left branch to operate as a pure LM only if the right branch lacks LM information.}
  \label{fig:speech_production}
\end{figure}

\section{Hybrid Attention-based Encoder-decoder}
\subsection{Formulation}
The Fig \ref{fig:speech_production} shows the high level architecture of standard AED model and HAED model we proposed. In the standard AED model, the decoder is responsible for generating both query for cross-attention and the hidden states based on previous tokens. To convert the decoder into an independent LM, it's necessary to segregate these two functions. Specifically, we must engineer the decoder to depend solely on previous predicted tokens and simultaneously eliminate its dependency on cross-attention. This second step is crucial, as we will demonstrate in the following section that the decoder can only function as an independent language model when the encoder and cross-attention are unaffected by predicted tokens. However, eliminating this dependency poses a significant challenge because the cross-attention must remain aware of the decoded tokens; otherwise, the cross-attention function will remain static through all decoding steps.

To achieve this, we replace $d_{u-1}$ with $h_{t}^{u-1}$ to remove attention's dependency on the decoder in equation \ref{attention}. Here, $h_{t}^{u-1}$ comes from encoder hidden states and the index $t$ and $u-1$ refers to the frame with highest CTC emission probability of label $y_{u-1}$, which we can simply obtain from the Forward–backward algorithm on CTC branch. This modification results in the updated equation: 
\begin{equation}\label{FAED attention}
  \begin{aligned}
\alpha_{u,t}=Softmax(Attention(h_{t}^{u-1}, h)))
\end{aligned}
\end{equation}
Without responsibility for generating cross-attention's query, the decoder in equation \ref{decoder} now becomes:
\begin{equation}\label{AED decoder}
  \begin{aligned}
d_u = Decoder(d_{u-1}, y_{u-1})=Decoder(y_{0:u-1})
\end{aligned}
\end{equation}
This equation has the exact same form as in an standard LM model. Finally, we can compute the output probability in equation \ref{output} for label $y_u$ now:
\begin{equation}\label{FAED output}
  \begin{aligned}
P(y_u|x, y_{0:u-1}) = Softmax(c_u + log\_softmax(d_u))
\end{aligned}
\end{equation}
It's worth noting that the decoder output is the log probability over the vocabulary to keep the decoder as an intact language model. Therefore, in theory, this internal language model could be replaced by any LM trained with the same vocabulary, such as LSTM and n-gram LMs.

In the training stage, the HAED is trained from scratch using the loss function defined:
\begin{equation}\label{faed_loss}
  \begin{aligned}
L = -lnP(y|x) - \lambda logP(y) - \beta L_{ctc}
\end{aligned}
\end{equation}
where the first term is the standard CE loss for AED models and the second term is the LM loss with cross entropy (CE). $\lambda$ is a hyper parameter to tune the effect of LM. We follow the recipe in \cite{kim2017joint} and incorporate a CTC model. In addition to faster convergence, this CTC model is also utilized to identify the frame with the highest CTC emission probability for tokens during training. This operation will not incur extra costs because we can determine the optimal CTC alignment simultaneously when calculating the CTC loss during training or when calculating the CTC prefix score during inference. Additionally, we can even accelerate this process in decoding by eliminating frames with a blank probability exceeding a certain threshold.

\subsection{Internal Language Model Score}
The intuition here is that the next token predicted by internal language model at label $u$ should only depend on the decoder as a function of previous $u-1$ labels, rather than the acoustic features. For example,referring to the high-level architecture in Fig \ref{fig:speech_production}, if right branch contains LM information, the left branch will not yield a pure LM score when they are combined to calculate the final score. In this section, we can prove that the internal language model is mostly proportional to the decoder in our HAED model: $P(y_u) \propto softmax(d_u)$. In our formulation, for each token $u$, the encoder and attention function $f(x, t^{u-1})$ encode input features X into $c_u$, and the decoder function $Decoder(y_{0:u-1})$ encode previous 0:u-1 label embeddings into $d_u$. Then both function are summed up to calculate the probability for token $y_u$ in equation \ref{FAED output} and the label posterior distribution is then obtained by normalizing the score functions across all labels in V:
\begin{equation}\label{posterior distribution}
  \begin{split}
P(y_u|x, y_{0:u-1}) = \frac{exp(S(y_u|x, y_{0:u-1}))}{\sum_{u' \in V}exp(S(y_{u'}|x, y_{0:u-1})) }
\end{split}
\end{equation}
where 
\begin{equation}\label{FAED score}
\begin{split}
S(y_u|x, y_{0:u-1}) = f(x, t^{u-1}) + log\_softmax(d_u)
\end{split}
\end{equation}
 from equation \ref{FAED output} where $f(x, t^{u-1})$ represents the operation which encodes x to h, and uses $h_t^{u-1}$ as query to do attention over h, and generates $c_u$, as we defined in equation \ref{FAED attention}.
 
As demonstrated in Proposition 1 in Appendix A \cite{variani2020hybrid}:
\begin{equation}\label{Proposition}
\begin{aligned}
\sum_xexp(S_{t,u}(y_u|x, y_{0:u-1})) \propto P(y_u, y_{0:u-1}) \\
\end{aligned}
\end{equation}
For HAED model, we also apply exponential function and marginalizing over x on equation \ref{FAED score}:
\begin{equation}\label{p3}
  \begin{aligned}
\sum_xexp(S(y_u|x, y_{0:u-1})) \\ = \sum_xexp(f(x, t^{u-1}) + log\_softmax(d_u)) \\
= softmax(d_u) \sum_xexp(f(x, t^{u-1}))
 \end{aligned}
\end{equation}
 The expression $\sum_xexp(f(x, t^{u-1}))$ is a constant only if $f$ is a function of $x$.  However, it is important to note that $f$ also takes $t^{u-1}$ as input and different $y_{u-1}$ may have distinct highest emission time index in CTC, resulting in varying values for $t_{u-1}$. In this situation, $y_{u-1}$ will have a impact on f, causing the encoder to rely on $y_{u-1}$, which is not desirable. To address this, we introduce an approximation by assuming that $y_{u-1}$ in all hypotheses have the same CTC highest emission index. We argue that, within the hypothesis space we are attempting to discriminate the most, the different $y_{u-1}$ usually corresponds to the same index. We make this assumption based on the observation that the most competitive hypotheses during decoding typically end at same CTC index in practice. By approximating $\sum_xexp(f(x, t^{u-1}))$ as a constant and we can have:

 \begin{equation}\label{output1}
  \begin{aligned}
\sum_xexp(S_u(y|x, y_{0:u-1})) \propto softmax(d_u)
 \end{aligned}
\end{equation}
 
Then given equation \ref{Proposition}, we can conclude that the internal LM is mostly learned by the decoder:
\begin{equation}\label{output1}
  \begin{aligned}
 softmax(d_u) \propto P(y_u, y_{0:u-1})
 \end{aligned}
\end{equation}
In other words, the internal LM is proportional to the posterior score of equation \ref{AED decoder} once the influence of $y_{0:u-1}$ on the encoder and attention function has been largely minimized. 

\begin{figure}[t]
  \centering
  \includegraphics[width=6cm]{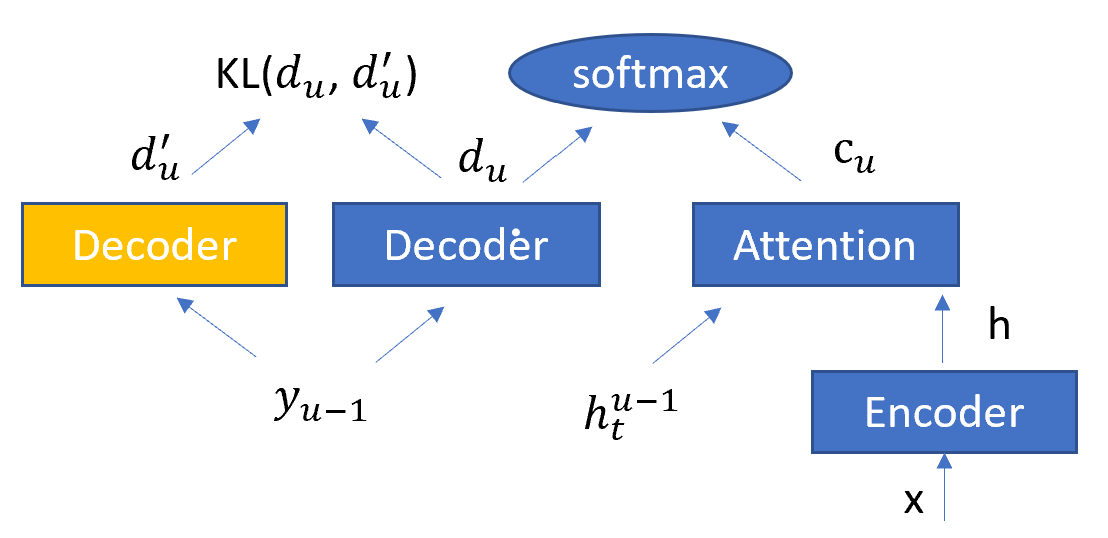}
  \caption{Text adaptation}
  \label{fig:adapt}
\end{figure}

\subsection{Text adaptation}
\label{ta}
During the text adaptation stage, we can utilize well-studied language model adaptation techniques to adapt the decoder to the target-domain text data since the decoder works as a language model. The most straightforward method is to directly fine-tune the decoder using the adaptation text data for a specified number of iterations. But this may degrade model performance on general domain. To avoid this, KL divergence between the decoder outputs of adapted model and baseline model is added during the adaptation as shown in figure \ref{fig:adapt}.

The adaption loss with KL divergence is:
\begin{equation}\label{output}
  \begin{aligned}
L = -lnP(y_{u}|y_{0:u-1})  - \alpha KL(d_u, d_u^{'})
\end{aligned}
\end{equation}
where $d_u$ is the output of decoder on adaptation text from the adapted model and $d_u^{'}$ is the output of decoder from the baseline model. $\alpha$ is the KL divergence weight to be tuned in the experiments.



\begin{table}[!ht]
    \caption{Statistics of test sets}
    \centering
    \begin{tabular}{l|l|l}
    \hline
        test set & adaptation data & testing data \\ \hline
        Conversation1 & 2k & 3k \\ \hline
        Boardcast & 4k & 6k \\ \hline
        Conversation2 & 17k & 2k \\ \hline
        Voice Assistant & 193k & 22k \\ \hline
        Lirispeech & 18740k & 210k \\ \hline
    \end{tabular}
    \label{stat}
\end{table}

\begin{table}[]
\centering
\begin{tabular}{l|l|l|l}
\hline
model                 & $\lambda$   & PPL & WER \\ \hline
AED                   &  -  &  -  &    10.56 \\ \hline
LM                    &  -  &  52.1  &   - \\ \hline
\multirow{4}{*}{HAED} & 0   &  292.6  &   10.86  \\ \cline{2-4} 
                      & 0.2 &  80.9  &  11.03   \\ \cline{2-4} 
                      & 0.5 &  71.3   &  10.90   \\ \cline{2-4} 
                      & 0.8 &  66.1  &   10.83  \\ \hline
\end{tabular}
\caption{WER(\%) on the general with different $\lambda$}
\vspace{-0.0cm}
\label{lm_weight}
\end{table}

\begin{table*}[th]
    \centering
    \begin{tabular}{l|c c c c c c c|c|c}
    \hline
       &Conversation1  &Conversation2 & Boardcast & Voice Assistant & Librispeech & Average \\
     \hline
      \multirow{3}{*}{}AED & 11.5 & 7.6 &  23.4 & 11.1 & 4.1 & 11.5 \\
      +SF & 11.4 & 5.2 & 19.5 & 8.6 & 3.8 & 9.7 \\
      +DR & 11.4 & 4.9 &  18.9 & \textbf {8.5} & 3.8 & 9.5 \\ 
    \hline
      \multirow{3}{*}{}HAED & 11.5 & 7.8  & 24.4 & 12.2 & 4.4 & 12.0\\
      +Adapt & \textbf {11.3} & 3.9  & 18.9 & 9.4 & 3.9 & 9.5 \\ 
      ++SF & 11.4 & \textbf {3.8} & \textbf {18.5} & 8.8 & \textbf{3.8} & \textbf{9.2}\\ 
      \hline
\end{tabular}
    \caption{WER(\%) on adaptation testing sets.}
    \vspace{-0.0cm}
    \label{adaptationset}
\end{table*}

\section{Experiments}
This section evaluates the effectiveness of the proposed methods for various adaptation tasks with different amounts of adaptation text data. The baseline AED model consists of an encoder network with 18 conformer layers, a decoder network with 6 transformer layers. The encoder and decoder have the embedding dimension 512, and number of HAEDs 8. The output label size is 4000 sentence pieces. For the HAED model, it has the same encoder structure and output label inventory as the baseline AED model. The decoder has two modules, one consists of 6 transformer layers without cross attention as LM and one consists of 6 layer with only cross attention. The acoustic feature utilized is the 80-dimension log Mel filter bank for every 10 ms speech from speech data sampled at 8K and 16K HZ. We train all the model for 225k steps with adamW optimizer on 8 V100 GPUs. During training and decoding, the CTC weight $\beta$ is set to 0.2 for both AED and HAED.

The training data contains 30 thousand (K) hours of transcribed Microsoft data, with personally identifiable information removed. The general testing set consists of 6.4 million words, covering various application scenarios such as dictation, conversation, far-field speech, and call centers. To evaluate the model's adaptability, we selected four real-world tasks with different sizes of adaptation text data and included Librispeech \cite{panayotov2015librispeech} sets for comparison. Table \ref{stat} provides details on the number of the words in each testing sets. The model training never sees the adaptation task data. 

For text adaption of HAED, we fine-tune the decoder for one sweep of adaptation data with constant learning rate of 5e-6 using standard CE loss and $\alpha=0.1$ for KL divergence loss. 



\begin{table}[th]
    \centering
    \begin{tabular}{l|c}
    \hline
          Model        & General set \\
     \hline
      Baseline AED      &    10.56    \\
    \hline
      HAED &  10.83 \\
      \quad w/o decoder & 12.11 \\
       \quad w/ ext. LM &  11.02\\

     \hline
    \end{tabular}
    \caption{WER(\%) on the general testing set.}
    \vspace{-0.0cm}
    \label{ab}
\end{table}

\subsection{Results on general testing sets}
The first experiment is to evaluate performance on the general testing set and results are shown in Table \ref{lm_weight}. The HAED models exhibit slight degradation in WER performance compared to the standard AED model. This is expected because the separation of encoder and decoder makes the joint optimization less effective. Then the HAED with different $\lambda$ as defined in Equation \ref{faed_loss} are also investigated. The higher LM weight of $\lambda$ results in decreasing PPL, but it did not impact much on the WER. This observation is similar to the one in HAT \cite{variani2020hybrid}. One possible explanation is that the encoder already learned a very good implicit internal LM and thus the explicit internal LM in the decoder is not as important as expected. As a reference, a Transformer-LM with same model structure as decoder trained on the text of the same training data results in a PPL of 52.1. In the following adaptation experiment, we adopt the HAED with $\lambda = 0.8$ as the seed model for LM adaptation on text data because it gives the lowest PPL and slighter better WER.


\subsection{Results on adaptation testing sets}
In this section, we investigate the language model adaptation of the HAED on the adaptation sets. The experiment results are reported in Table \ref{adaptationset}. The results in the “AED” row are the WER for standard AED model before adaptation. Since it is not possible to fine-tune the decoder of the AED model on text, we use the shallow-fusion \cite{gulcehre2015using} technique, where we used a 5-gram word-piece level n-gram LM built on the adaptation text data. The interpolation weight is set to 0.1, which is found to perform well across all testing sets. The results are presented in the “+SF” column. Additionally, in the "+DR" column, we explore the use of the density ratio (DR) approach \cite{mcdermott2019density} for adaptation. We build a 5-gram word-piece level n-gram LM in source domain with a weight of 0.1 and 5-gram word-piece level n-gram LM in target domain with a weight of 0.1. These weights are tuned to achieve optimal performance across all sets.  

Similar to the results on the general test set, the performance of the HAED is worse than the standard AED without adaptation. However, by using adaptation approach described in section \ref{ta} on the target-domain text data, significant WER reductions can be achieved, as shown in the "+adapt" column. This results in a relative 20\% WER improvement relatively on the adaptation test sets. In comparison, using the shallow fusion approach results in a 16\% WER reduction relatively, and using the density ratio approach results in a 17\% relative WER reduction on the standard AED model. We also found the adapted HAED model can be further improved using the shallow fusion \cite{gulcehre2015using}, as indicated in the "++SF" column. This combined approach results in a 23\% relative WER reduction, achieving the best performance on most test sets.

\subsection{Ablation study}
In our previous experiments, we noticed that the improvement of LM on decoder network does not yield lower WER. Therefore, we decided to investigate the effect of the decoder in HAED. We first trained a HAED model without the decoder network, the output in equation \ref{FAED output} now became: $P(y_u|x, y_{u-1})=Softmax(c_u)$. As shown in Table \ref{ab}, the performance degrades a lot. The decoder network might not behave as a real language model but it's still a significant component. And we also trained a HAED model with decoder initialized by an external LM. This external LM has the same structure as the decoder and was trained on text data containing about 1.5 billion words, including the transcription of the 30k training data mentioned earlier. The external LM parameters is updated together with the model during training. Surprisingly, we found that initializing the model with the external LM did not yield any improvements over the baseline. We hypothesized that strong internal language model may not be essential, as the encoder likely captures most of the information already.

\section{Conclusion}
In this work, we propose a novel hybrid attention-based encoder-decoder model that enables efficient text adaptation in an end-to-end speech recognition system. Our approach successfully separates the acoustic model and language model in the attention-based encoder-decoder system. Such separation allows us to easily adapt the decoder to text-only data, similar to conventional hybrid speech recognition systems. Our experiments on multiple sets demonstrated the effectiveness of the proposed methods. Compared with the standard AED model, our method can substantially reduce the word error rate with adaptation text data. For future work, we aim to investigate using a large language model as the decoder or training the model with both audio-text pair data and text-only data.

\bibliographystyle{IEEEbib}
\bibliography{strings,refs}

\end{document}